\documentclass[journal=nalefd,manuscript=article,numbers]{achemso}

\DeclareUnicodeCharacter{2212}{-}
\usepackage{dcolumn}
\usepackage{bm}

\usepackage{amsmath}
\usepackage{amssymb} 
\usepackage{bm}  
\usepackage{soul}
\usepackage{mathtools}
\usepackage{upgreek}
\usepackage{array}
\usepackage{setspace}   
\usepackage{tabulary}
\usepackage{ulem}
\usepackage{xcolor}
\newcolumntype{P}[1]{>{\centering\arraybackslash}p{#1}}
\usepackage{caption}
\usepackage{natbib}

\title{Room Temperature Exciton-Polariton Condensation in Silicon Metasurfaces Emerging from Bound States in the Continuum}

\author{Anton Matthijs Berghuis}
\email{a.m.berghuis@tue.nl}
\affiliation{Department of Applied Physics and Science Education\\ and Eindhoven Hendrik Casimir Institute, Eindhoven University of Technology,\\
P.O. Box 513, 5600 MB Eindhoven, The Netherlands.}
\affiliation{Institute for Complex Molecular Systems-ICMS, Eindhoven University of Technology, P.O. Box 513, 5612 AJ, Eindhoven, The Netherlands}
\author{Gabriel W. Castellanos}
\affiliation{Department of Applied Physics and Science Education and Eindhoven Hendrik Casimir Institute, Eindhoven University of Technology,\\
P.O. Box 513, 5600 MB Eindhoven, The Netherlands.\\
}
\affiliation{Institute for Complex Molecular Systems-ICMS, Eindhoven University of Technology, P.O. Box 513, 5612 AJ, Eindhoven, The Netherlands}

\author{Shunsuke Murai}
\affiliation[kyoto]
{Department of Material Chemistry, Graduate School of Engineering, Kyoto University, Katsura, Nishikyo, 6158510, Kyoto, Japan
}

\author{Jose Luis Pura}
\affiliation{Instituto de Estructura de la Materia (IEM-CSIC), \\
Consejo Superior de Investigaciones Científicas, \\
Serrano 121, 28006 Madrid, Spain.\\
}

\author{Diego R. Abujetas}
\affiliation{Physics Department, \\
Fribourg University, \\
Chemin de Musée 3, Fribourg 1700, Switzerland.\\
}
\author{Erik van Heijst}
\affiliation{Department of Applied Physics and Science Education and Eindhoven Hendrik Casimir Institute, Eindhoven University of Technology,\\
P.O. Box 513, 5600 MB Eindhoven, The Netherlands.\\
}
\affiliation{Institute for Complex Molecular Systems-ICMS, Eindhoven University of Technology, P.O. Box 513, 5612 AJ, Eindhoven, The Netherlands}

\author{Mohammad Ramezani}
\affiliation{Department of Applied Physics and Science Education and Eindhoven Hendrik Casimir Institute, Eindhoven University of Technology,\\
P.O. Box 513, 5600 MB Eindhoven, The Netherlands.\\
}
\affiliation{Institute for Complex Molecular Systems-ICMS, Eindhoven University of Technology, P.O. Box 513, 5612 AJ, Eindhoven, The Netherlands}

\author{Jos\'e A. S\'anchez-Gil}
\affiliation{Instituto de Estructura de la Materia (IEM-CSIC), \\
Consejo Superior de Investigaciones Científicas, \\
Serrano 121, 28006 Madrid, Spain.\\
}

\author{Jaime G\'omez Rivas}
\email{j.gomez.rivas@tue.nl}
\affiliation{Department of Applied Physics and Science Education and Eindhoven Hendrik Casimir Institute, Eindhoven University of Technology,\\
P.O. Box 513, 5600 MB Eindhoven, The Netherlands.\\
}
\affiliation{Institute for Complex Molecular Systems-ICMS, Eindhoven University of Technology, P.O. Box 513, 5612 AJ, Eindhoven, The Netherlands}

\begin{document}

\begin{abstract}
We show the first experimental demonstration of room-temperature exciton-polariton (EP) condensation from a bound state in the continuum (BIC). This demonstration is achieved by strongly coupling stable excitons in an organic perylene dye with the extremely long-lived BIC in a dielectric metasurface of silicon nanoparticles. The long lifetime of the BIC, mainly due to the suppression of radiation leakage, allows for EP thermalization to the ground state before decaying. This property results in a condensation threshold of less than 5 $\upmu \text{J cm}^{-2}$, one order of magnitude lower that the lasing threshold reported in similar systems in the weak coupling limit.
\end{abstract}
\maketitle
\setstretch{1.65}


\section*{Introduction}
A Bose-Einstein condensate (BEC) is a system of bosonic (quasi-)particles that have undergone thermalization to occupy the ground state. The phase transition to form a BEC occurs at a critical temperature that depends on the effective mass of the bosons.
Condensation of exciton-polaritons (EPs) has been widely investigated over the last decades~\cite{Kasprzak2006Bose-EinsteinPolaritons,Kena-Cohen2010Room-temperatureMicrocavity,Byrnes2014Exciton-polaritonCondensates,Plumhof2014Room-temperaturePolymer,Ramezani2017Plasmon-exciton-polaritonLasing,Ardizzone2022PolaritonContinuum,Riminucci2022NanostructuredContinuum}. EPs are bosonic quasi-particles that result from the strong coupling of photons in an optical cavity and excitons in a semiconductor~\cite{Weisbuch1992,Lidzey1998}. Due to their low effective mass, EPs can form condensates at high temperatures and even at room temperature when the EP binding energy is sufficiently large. EPs in a condensate occupy the same quantum state. Therefore, when EPs decay emitting radiation, this radiation may leak from the cavity producing coherent emission, known as polariton lasing. Polariton lasers do not require population inversion, which potentially enables lasing at much lower thresholds than conventional lasers\cite{Byrnes2014Exciton-polaritonCondensates}. Therefore, EP condensates offer a promising alternative to achieve continuous wave and electrically driven laser-like emission from solid-state organic devices. \\
\\
The first demonstration of EP condensation was realized with inorganic quantum wells in a Fabry-Perot microcavity at cryogenic temperatures~\cite{Kasprzak2006Bose-EinsteinPolaritons}. Subsequent research has shown that strong light-matter coupling with excitons in organic semiconductors, which have much larger binding energies than excitons in inorganic materials, can also lead to EP condensation at room temperature\cite{Kena-Cohen2010Room-temperatureMicrocavity,Plumhof2014Room-temperaturePolymer}. While most polariton condensation experiments have been conducted in Fabry-Perot microcavities, this phenomenon has also been observed in optical metasurfaces supporting the so-called surface lattice resonances (SLRs)\cite{Ramezani2017Plasmon-exciton-polaritonLasing,Hakala2018Bose-EinsteinLattice}. The latter offers the advantage of easy fabrication over large areas and the possible application in integrated photonics. In contrast to a gas of atoms, the original platform for BEC\cite{Anderson1995ObservationVapor}, exciton polaritons have very short lifetimes.
These short lifetimes limit the build-up of the EP density at the ground state, which results in an increased threshold for condensation. Consequently, EP condensation requires powerful laser systems to produce a sufficiently high number of excitons and reach the threshold, which makes polariton lasing unsuitable for most applications. \\
\\
In this manuscript, we demonstrate low threshold EP condensation by reducing significantly the losses in an all-dielectric cavity formed by a silicon (Si) metasurface, thus increasing the EP lifetime. Recent efforts have succeeded to reduce the condensation thresholds by replacing metallic metasurfaces supporting plasmonic SLRs with low-loss dielectric metasurfaces supporting Mie-SLRs~\cite{Castellanos2023NonEquilibriumMetasurfaces}. Due to the high Q-factors of the SLRs between 400 and 700 nm, partly due to the reduction of material losses, the condensation threshold was reduced significantly. Here, we explore the limits of organic EP condensation by also suppressing radiation losses. This suppression is achieved by coupling excitons to symmetry-protected bound states in the continuum (BICs) supported by the array of Si Mie-resonators. BICs are optical modes with infinitely long lifetimes in lossless surfaces due to the cancellation of the radiation leakage. This suppression of radiation leakage is imposed by the symmetry mismatch between the mode profiles at the surface and those of the radiation continuum, which results in a vanishing of the overlap integral\cite{Marinica2008BoundPhotonics}. Zhen et al. showed that these modes are associated with a topological charge, making them robust to perturbations and visible in the far-field as a polarization vortex. \cite{Zhen2014TopologicalContinuum}. By the strong coupling of excitons to BICs in a dielectric metasurface, we reduce the EP condensation threshold to 5 $\upmu \text{J cm}^{-2}$.

The lack of radiative losses in BICs, despite the fact that these are modes in the radiation continuum, has recently attracted significant research interest in various fields. BICs are a promising platform for photon lasing, which has been demonstrated for various gain media, such as quantum wells\cite{Noda2001PolarizationDesign,Kodigala2017LasingContinuum,Hwang2021Ultralow-thresholdContinuum}, quantum dots \cite{Guan2020QuantumPatterns, Wu2020Room-TemperatureContinuum,Wu2021BoundLasing}, organic materials, \cite{Hakala2017LasingLattice,Yang2021Low-ThresholdMetasurfaces,Mohamed2022ControllingContinuum,Zhai2022MultimodeContinuum} or with the semiconducting metasurface itself as gain medium\cite{Ha2018DirectionalArrays}. 
These systems are in the weak light-matter coupling regime, thus corresponding to conventional photon lasers where population inversion is required to achieve a net gain and lasing action. However, in the strong light matter-coupling regime, it is possible to reach polariton condensation and coherent emission without population inversion and at potentially lower thresholds. 
Strong light-matter coupling with BICs has been observed for several systems \cite{Koshelev2018StrongContinuum,Lu2020EngineeringPoints,Kravtsov2020NonlinearContinuum, Dang2022RealizationMetasurface,Al-Ani2022StrongMetasurfaces}, but only very recently polariton condensation has been reported in a system of GaAs quantum wells at cryogenic temperatures\cite{Ardizzone2022PolaritonContinuum,Riminucci2022NanostructuredContinuum}. The low exciton binding energies in inorganic semiconductors, typically below the thermal energy at room temperature, makes low temperatures necessary. To overcome this limitation, we design dielectric metasurfaces supporting BICs and couple them to organic molecules. We reach the strong light-matter coupling regime and achieve EP condensation from a BIC at room temperature and low thresholds. This result sets an important step forward towards the realization of electrically driven coherent emission from organic systems.

\section{Results}
\subsection{Bare Si metasurface}
\begin{figure}[H] 
  \begin{center}
    \includegraphics[width=.45\textwidth]{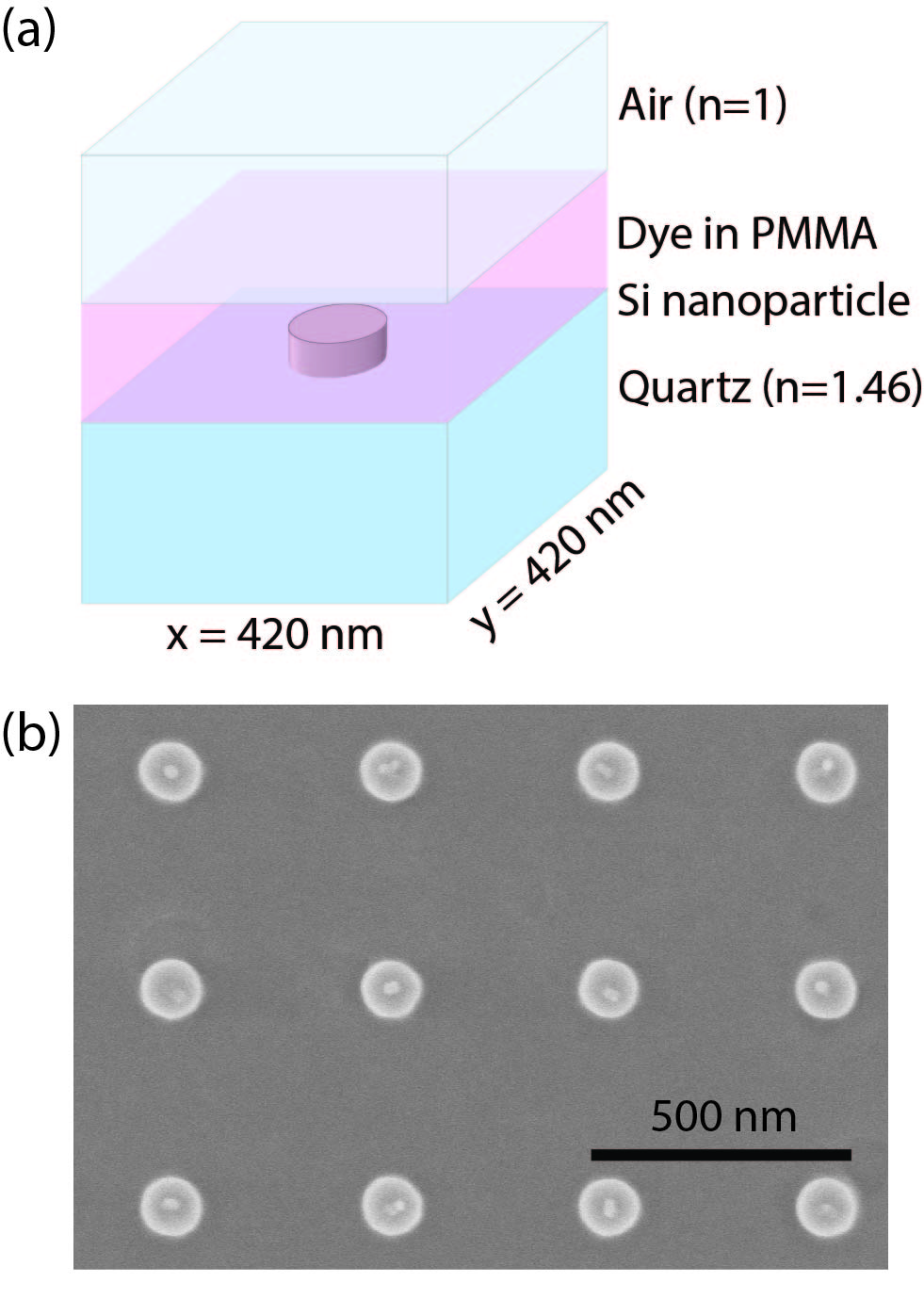}
    \end{center}
  \caption{\setstretch{1.65} (a) Schematic overview of a unit cell of the Si metasurface, covered with a 200 nm layer of perylene dye in PMMA. (b) Top view SEM image of the Si nanoparticle array.}
  \label{fgr:Figure1}
\end{figure}

\begin{figure}[H] 
  \begin{center}
    \includegraphics[width=.85\textwidth]{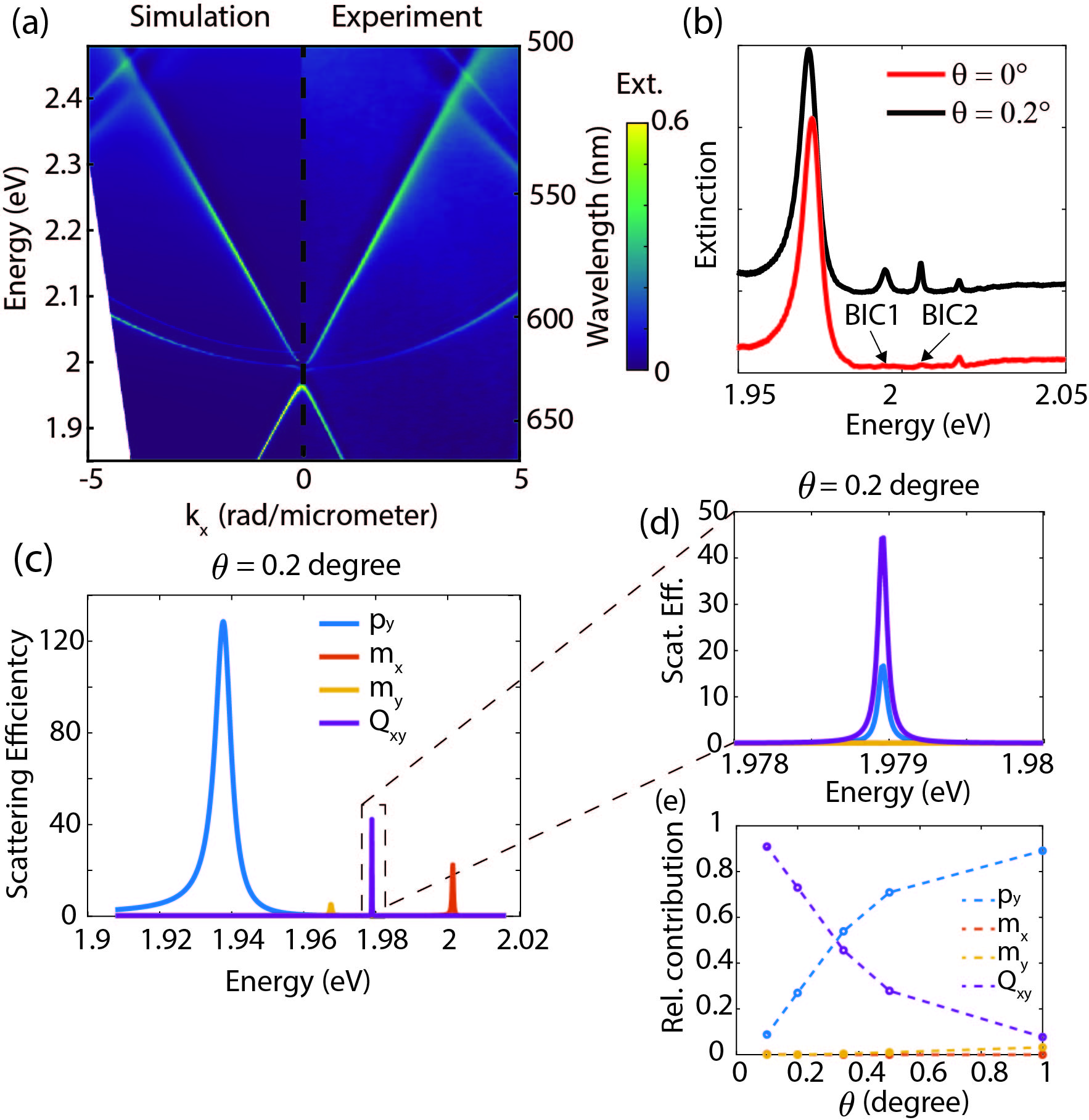}
    \end{center}
  \caption{\setstretch{1.65} (a) Simulated (left panel) and experimental (right panel) extinction (1-transmission) spectra of the silicon metasurface with a 230 nm thick layer of polystyrene (n=1.59) on top as a function of the parallel component of the wave vector of the incident wave, i.e., $\frac{2\pi}{\lambda} \sin\theta$, with $\theta$ the angle of incidence. (b) Measured extinction spectra at $\theta=0^\circ$ and $\theta=0.2^\circ$. The energies of the two BICs at normal incidence are indicated. (c) Multipolar decomposition of the resonances for a plane wave incident at 0.2$^\circ$. (d) Magnified view of (c). (e) Angle dependent contributions of the different multipoles for BIC 2 (i.e. the mode that becomes a BIC at k=0 at 1.979 eV).}
  \label{fgr:Figure2_bareArray}
\end{figure}

The investigated sample, consisting of a square array of Si nanodisks (heigth h=90 nm, diameter d=90 nm, lattice constant P=420 nm) on top of a quartz substrate, is represented schematically in Fig. \ref{fgr:Figure1}a. Arrays of poly-crystalline Si nanodisks were fabricated using electron beam lithography as described in the methods section and illustrated by the scanning electron microscopy (SEM) image shown in Fig. \ref{fgr:Figure1}b. The poly-crystalline nature of the particles results in lower material losses compared to amorphous silicon, which is essential for a low condensation threshold.\\
\\
When the array is embedded in a medium with a homogeneous refractive index, the Mie-resonances of the individual nanoparticles can couple radiatively through in-plane diffractive orders, resulting in a forward and backward propagating transverse electric (TE) Mie SLR (TE-SLR) and a degenerate transverse magnetic TM-SLR (see Supporting Information (SI), section S1)~\cite{Castellanos2020Exciton-PolaritonsMetasurfaces}. 
When the Si metasurface is covered with a higher refractive index film (polystyrene film with a thickness of 230 nm and refractive index n=1.59), the SLRs can couple with the quasi-guided modes in the film~\cite{Tikhodeev2002QuasiguidedSlabs,Murai2013HybridWaveguides}. This coupling allows for the efficient excitation of additional modes as it is apparent in the angle dependent extinction measurements and simulations using the Rigorous Coupled Wave Analysis (RCWA) method shown in Fig.~\ref{fgr:Figure2_bareArray}a (see the Methods section for details about measurements and simulations). The two TE-SLRs with an anti-crossing at normal incidence\cite{Barnes1996} are the dominant modes in this angle dependent dispersion, but additionally two weaker 'parabolic' modes are visible. Interestingly, two of these modes become extremely narrow approaching normal incidence and vanish completely at normal incidence, as can be clearly appreciated in the extinction measurements at the angles of incidence of $\theta=0^\circ$ and $0.2^\circ$ shown in Fig.~\ref{fgr:Figure2_bareArray}b. This characteristic indicates that these modes are symmetry-protected BICs. \\
\\
To understand the character of these BICs, we have performed a multipolar decomposition of the scattering efficiency by the resonances at different energies, for the array covered with a dye doped layer. Using this method we retrieve the character of the resonances in terms of the electric dipolar, magnetic dipolar and electric quadrupolar modes (see Methods section for details). The result of the decomposition for an incident plane wave at an angle of 0.2$^\circ$ is shown in Fig.~\ref{fgr:Figure2_bareArray}c. The mode at $\sim$1.94 eV (blue peak in Fig.~\ref{fgr:Figure2_bareArray}c) corresponds to an electric dipolar mode along the y-direction, while the mode at $\sim$2 eV (red peak) corresponds to a magnetic dipolar mode along the x-direction. The two modes that become a BIC at normal incidence at $\sim$ 1.965 eV (yellow peak, BIC 1) and 1.979 eV (purple peak, BIC 2) have a dominant magnetic dipolar and electric quadrupolar character, respectively. The decomposition of BIC 2 is shown in more detail in Fig.~\ref{fgr:Figure2_bareArray}d (for details about the decomposition of the other modes see section S2 in the SI). Besides the quadrupolar contribution, there is a dipolar contribution as well, leading to the excitation of the mode at non-zero angles. Figure ~\ref{fgr:Figure2_bareArray}e shows how the character of BIC 2 changes from dipolar at large angles to quadrupolar towards normal incidence. This explains why at normal incidence all radiation losses are suppressed as a pure quadrupolar mode in an infinite lattice can neither be excited
from the far field, nor radiate into the far field.
The full suppression of radiation losses of the BICs will be exploited for low-threshold polariton condensation when the resonances in the metasurface are coupled to excitons in a film of organic molecules.\\

\subsection{Strong coupling to excitons}
\begin{figure}[H] 
  \begin{center}
    \includegraphics[width=.95\textwidth]{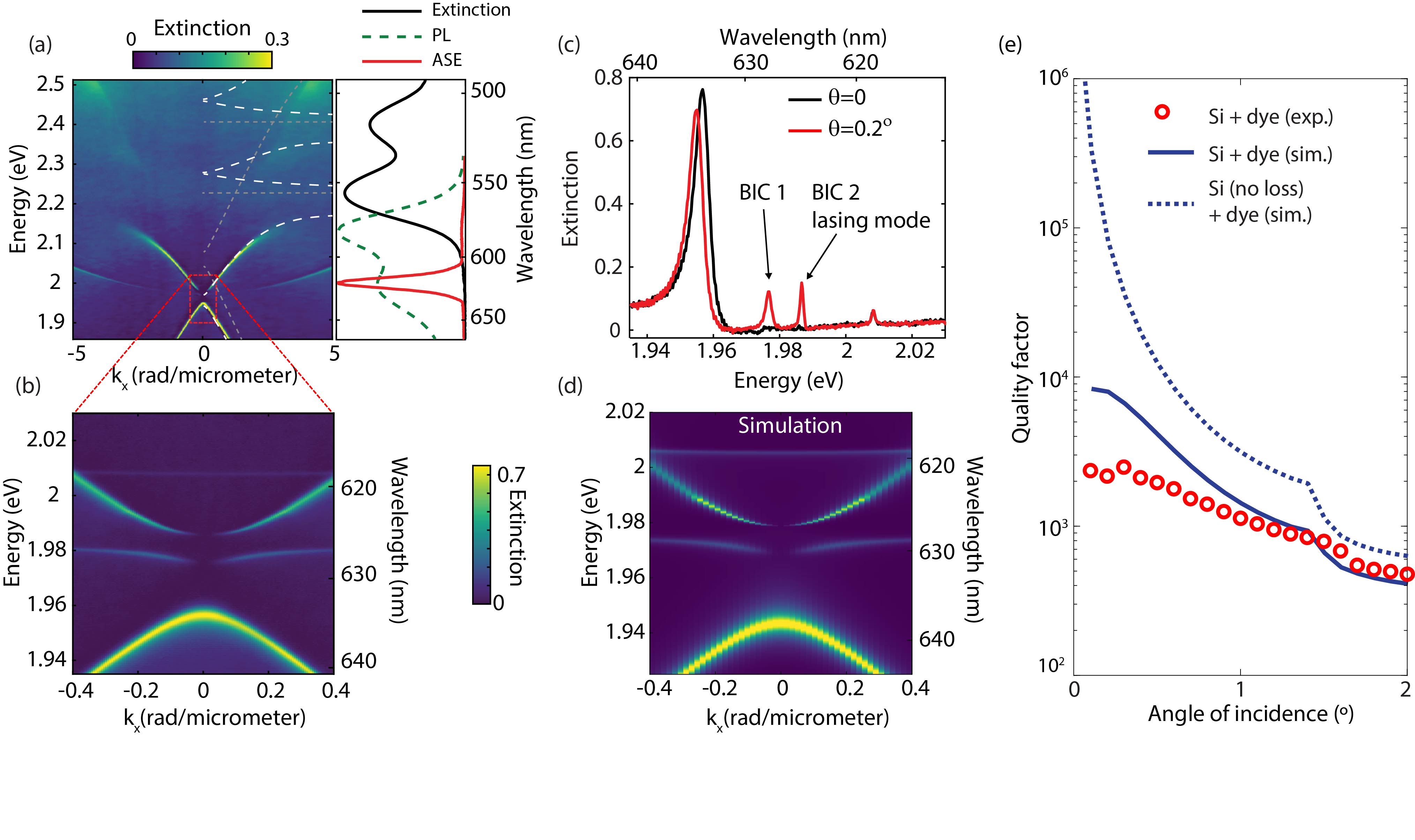}
    \end{center}
  \caption{\setstretch{1.65} (a) Optical extinction of the array of Si nanodisks as a function of in-plane momentum (k$_x$) measured with a Fourier microscope. The gray dashed lines indicate the energies of the bare SLRs and the exciton energies. The white dashed curves are the EP dispersion resulting from the coupling of the SLRs with the excitons, as calculated with a 4-state coupled oscillator model. The right panel of (a) shows the extinction, photoluminescence, and amplified spontaneous emission of the bare layer of perylene dye in PMMA. (b) Magnified view of (a), where two BIC modes are dark at normal incidence. (c) Extinction spectra measured at $\theta=0^\circ$ and $0.2^\circ$ showing two narrow modes for $\theta=0.2^\circ$. (e) Quality factor of the modes turning into  BIC 2, as a function of the angle of incidence. The red circles indicate the Q-factor calculated from the measured width of the resonance, the blue curve shows the simulated Q-factor for the resonance for silicon nanoparticles with realistic losses, and the blue-dashed curve shows the Q-factor for Si without losses.
  }
  \label{fgr:Figure3_Extinction}
\end{figure}

To investigate the formation and condensation of EPs, we remove the transparent polystyrene film on top of the array and spin coat a solution of 32 wt\% perylene dye ([N, N'-Bis(2,6-diisopropylphenyl)-1,7- and -1,6-bis (2,6-diisopropylphenoxy)-perylene-3,4:9,10-tetracarboximide]) in poly(methyl methacrylate) (PMMA) to form a 200 nm thick layer. The bare molecules show two exciton peaks at 2.24 and 2.41 eV, corresponding to the electronic transition and its first vibronic replica, respectively. These peaks are visible in the extinction spectrum plotted with the black curve in the right panel of Fig. \ref{fgr:Figure3_Extinction}a. The green dashed curve in the same panel shows the normalized photo-luminescence (PL) emission spectrum from the layer when excited by a 400 nm laser with a repetition rate of 1 kHz at low fluence, and the red curve corresponds to the normalized amplified spontaneous emission (ASE) spectrum measured at a fluence of $>$500 $\upmu \text{J cm}^{-2}$.\\
\\

A high concentration of molecules in the film is required to achieve collective strong light-matter coupling, as the coupling strength scales with the square root of the number of dye molecules in the mode volume, i.e., $g=g_0\sqrt{N}$ where $g_0$ is the coupling strength for a single molecule in the cavity, N the number of molecules and $g$ the collective coupling strength~\cite{Tavis1969}. The collective strong coupling of the molecules and the lattice resonances is manifested in the momentum-resolved transmission measurement as an anti-crossing of the exciton resonance and the TE-SLRs, as shown in Fig.~\ref{fgr:Figure3_Extinction}a. The dispersion of the resulting EPs can be fitted by a four-level coupled oscillator model incorporating the forward and backward propagating TE-SLRs and two exciton resonances~\cite{Rodriguez2013} (details of the coupled oscillator model are given in section S3 of the supplementary information). The energies of these exciton transitions and the dispersion of the bare cavity modes are plotted with the gray dotted lines in Fig. \ref{fgr:Figure3_Extinction}a (for $k_x>0$, and the resulting coupled resonances with the white-dashed curves. At normal incidence ($k_{x}$=0), the EPs energies overlap with the ASE, indicating that these polaritons are in the region with the highest gain in the film.\cite{Samuel2007OrganicLasers}\\
\\
We zoom in further in the dispersion measurements to the region around $k_{x}$=0 by using a low numerical aperture (NA) objective (Nikon 10x, 0.3 NA) and a finer grating in the spectrometer (600 lines/mm). We observe that, similar to the array covered with polystyrene, two of the four modes show an increasing quality factor towards normal incidence, eventually becoming dark at $k_{x}$=0 (see Fig.~\ref{fgr:Figure3_Extinction}b). The measured dispersion matches excellently with the RCWA simulations (Fig.~\ref{fgr:Figure3_Extinction}d).  
The narrow linewidth at small angles of incidence becomes especially clear in the spectra taken at $\theta=0.2^\circ$ and normal incidence, as plotted with the red and black curves in Fig.~\ref{fgr:Figure3_Extinction}c. From the spectrum at $\theta=0.2^\circ$, there is a difference in the linewidth between the mode at 1.977 eV (BIC 1, magnetic dipolar character) and the mode at 1.987 eV (BIC 2, quadrupolar character), where the latter clearly has a higher Q-factor (See SI, section S4 for a detailed analysis of the fields and Q-factors of both modes). 

To investigate the properties of BIC 2 in more detail, we fit the resonance width with a Fano lineshape\cite{Miroshnichenko2010FanoStructures,Limonov2017FanoPhotonics} 
\begin{equation}
\sigma(\lambda)=\alpha sin^2\delta \frac{(cot\delta+\Omega)^2}{1+\Omega^2},
\label{eq:Fano}
\end{equation}
where $\Omega=2\frac{(\lambda-\lambda_0)}{\Gamma}$, $\lambda_0$ the resonance wavelength, $\Gamma$ the linewidth, $\delta$ the Fano parameter and $\alpha$ the amplitude of the resonance. From the linewidth and resonance wavelength, we calculate the Q-factor as $Q=\lambda_0/\Gamma$. The resulting Q-factor as a function of the angle of incidence obtained from the experimental transmission spectra is plotted with the red circles in Fig.~\ref{fgr:Figure3_Extinction}e. The Q-factor increases towards smaller angles until it reaches 
a Q-factor of around 2200. As expected for any real system with unavoidable imperfections and a limited size, the Q-factor does not diverge. However, an experimental Q-factor of 2200 is among the highest reported for Mie metasurfaces in the visible range~\cite{Bin-Alam2021Ultra-high-QMetasurfaces}.

We compare the experimental Q-factor with the Q-factor extracted from RCWA simulations by fitting the simulated dispersions with the Fano profile (Eq.~\ref{eq:Fano}). The blue curve in Fig.~\ref{fgr:Figure3_Extinction}e shows the calculated Q-factors for the simulated array of Si nanorods with a permittivity accounting for the losses in the silicon nanoparticles. The simulated Q-factor follows a similar trend as the experimental one but reaches a value of approximately $10^4$ at small angles. At this point, material losses limit further increase of the Q-factor. If we set these losses equal to zero in the RCWA simulations, the Q-factor indeed diverges when approaching $k_x=0$, as expected for a symmetry protected BIC (dotted curve in Fig.~\ref{fgr:Figure3_Extinction}e). 
\\

\begin{figure}
  \begin{center}
    \includegraphics[width=.9\textwidth]{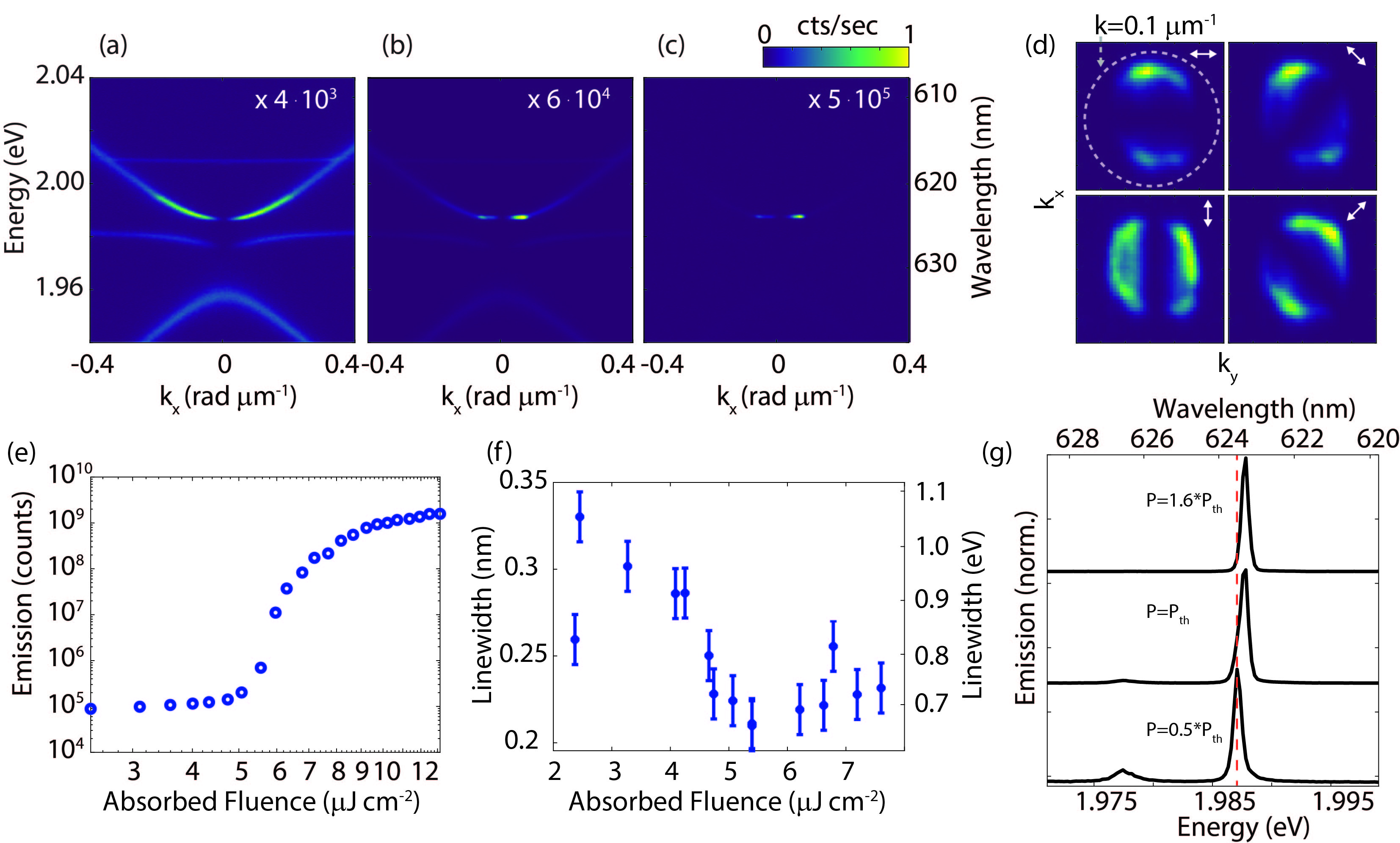}
    \end{center}
  \caption{\setstretch{1.65} Emission spectra as a function of in-plane momentum for different pump fluences: (a) below threshold, (b) on threshold, and (c) above threshold. (d) Coherent emission for an excitation power of 1.2$P_{th}$ as function of $k_x$ and $k_y$, evidencing the vortex behavior. (e) Emission intensity versus pump fluence integrated over $k_y$ and $k_x$ from -0.15 to 0.15 rad $\upmu \text{m}^{-1}$. (f) Full-width at half-maximum (FWHM) of the emission as a function of pump fluence. (g) Emission spectra below and above the threshold, showing the blue shift of the emission peak as the pump fluence is increased.}
  \label{fgr:Figure4_Lasing}
\end{figure}

If we pump the system non-resonantly with an amplified laser system (Vitarra, $\lambda=$400 nm, rep. rate = 1 kHz, pulse duration $\sim$ 150 fs) at low fluences, we can resolve the $k_x$ and energy-dependent fluorescent emission of the sample (See Fig.~\ref{fgr:Figure4_Lasing}a). This emission is clearly dominated by the decay into the different modes and the outcoupling into the far-field by the metasurface, with the strongest emission from the lower polariton band (BIC 2). This process alters completely the emission spectrum of the bare molecules. When the pump fluence is increased to approximately $P_{th}=5$ $\upmu$J $\text{cm}^{-2}$, the emission mostly originates from two points in k-space around 1.987 eV, indicating that the polariton lasing threshold is reached (Fig.~\ref{fgr:Figure4_Lasing}b). At higher pump fluences of $1.2 P_{th}$ the emission is fully dominated by this coherent emission (Fig~\ref{fgr:Figure4_Lasing}c). Since the BIC itself is non-radiative at $k_x=0$, a vortex beam in reciprocal space is formed with the light being emitted from angles slightly off normal incidence. Further polarized analysis of the emission as a function of $k_x$ and $k_y$, shows the vortex behavior of the emission(Fig.~\ref{fgr:Figure4_Lasing}d), experimentally evidencing the topological character of the BIC~\cite{Zhen2014TopologicalContinuum} and its quadrupolar nature~\cite{Wu2021BoundLasing,Zhai2022MultimodeContinuum}.

We monitor the non-linear emission by plotting in Fig.~\ref{fgr:Figure4_Lasing}e the integrated emission over a solid angle from k$_{x}$,k$_{y}$=-0.15 to k$_{x}$,k$_{y}$=0.15 $\upmu \text{m}^{-1}$, evidencing the condensation threshold at $P_{th}=5$ $\upmu$J $\text{cm}^{-2}$. We further confirm the transition from spontaneous emission to polariton lasing and the formation of the EP condensate by measuring the reduction in the linewidth, as plotted in Fig. \ref{fgr:Figure4_Lasing}f for $k_x=0.05 \upmu \text{m}^{-1}$. This linewidth reduction is a clear indication of the increased temporal coherence of the condensate. Finally, Fig.~\ref{fgr:Figure4_Lasing}g illustrates the blue shift of the emission peak as the pump fluence increases, characteristic of condensates~\cite{Plumhof2014Room-temperaturePolymer,DeGiorgi2018InteractionCondensate,Yagafarov2020MechanismsCondensates}. 

We note that the observed condensation threshold of 5 $\upmu \text{J cm}^{-2}$ is 40\% lower than previously reported values from SLRs in similar arrays of Si nanoparticles,~\cite{Castellanos2023NonEquilibriumMetasurfaces} being the lowest reported condensation threshold at room temperature. This threshold is comparable to the recent results at cryogenic temperatures in inorganic semiconductors, illustrating the potential of BICs in metasurfaces for room-temperature condensation in organic systems at ultra-low thresholds.\\
\\

\section{Conclusions}
In conclusion, we have demonstrated room-temperature exciton-polariton condensation from a BIC in a metasurface of Si nanoparticles resulting from SLRs. The lack of radiative losses and the low material losses of Si results in very high-quality factors of the polariton mode. The condensation threshold of 5 $\upmu \text{J cm}^{-2}$ is the lowest reported threshold for an organic small molecule polariton condensate and comparable to values measured at low temperatures in inorganic systems. These results set an important step towards the realization of room-temperature electrically pumped organic polariton lasers, which require low thresholds.  

\section{Methods}
\subsection{Fabrication of the Silicon metasurface}
Polycrystalline Si thin films with a thickness of 90 nm were grown on a synthetic silica glass substrate by low-pressure chemical vapor deposition using SiH$_4$ gas as a source of Si. A resist (NEB22A2, Sumitomo) was cast on the Si film and exposed to electron-beam lithography, followed by development to make nanoparticle arrays of resist on the Si film. The Si film was vertically etched using a selective dry etching (Bosch process) with SF$_6$ and C$_4$F$_8$ gases, and the resist residue was etched away by oxygen dry etching. The fabricated array covered and area of 2.5 × 2.5 mm$^2$

\subsection{Fourier microscopy measurements}
To map the energy and momentum resolved extinction of the metasufaces, we use a Fourier microscope in transmission mode. 
The excitation objective (40x, 0.6NA) focuses a white light onto the sample. The transmitted light is collected by a 60x, 0.7 NA or 10x, 0.3 NA objective. The back focal plane of this objective is imaged on the spectrometer slit (Princeton Instruments SP2300) using 2 lenses in a 4f configuration. The spectrometer transmits light along one of the principal axis of the metasurface ($k_x$ or $k_y$) and disperses light on the CCD camera (Princeton Instruments ProEM:512) using a 150 or a 600 lines/mm grating. For fluorescence/lasing measurements, a 400 nm laser at 1 kHz is used. This beam is generated by frequency doubling the output of a Ti:sapphire regenerative amplifier (Coherent Astrella), $\lambda$= 800 nm with a pulse duration of 150 fs.

\subsection{Rigorous coupled-wave analysis simulations}
For the simulated dispersion of the dielectric metasurfaces we used rigorous coupled-wave analysis (RCWA).  RCWA is a semi-analytical method that solves Maxwell's equations in Fourier space and offers great simulation speed for dielectric structures. We developed a homemade code based on Refs. [\citenum{Moharam1995FormulationGratings,Rumpf2006DesignBehavior}]. In addition to conventional RCWA the Normal Vector Method \cite{Schuster2007NormalGratings} was applied to increase convergence of the simulations. This allowed the use of 11 x 11 spatial harmonics to accurately describe the unit cell for the calculation of the transmittance of a plane wave through the sample. The permittivity values of Si were taken from literature \cite{Palik} and the permittivity of the dye-doped layer were obtained from ellipsometry measurements.

\subsection{COMSOL simulations}
The multipole decomposition of the observed modes was performed with the Electromagnetic Waves in Frequency Domain module of COMSOL Multiphysics. The simulation was done for the array covered with the rylene dye molecules in PMMA, using the same material properties and configuration of the RCWA simulations, we calculated the polarization $\mathbf{P}$ induced in the system by external plane waves impinging at different angles with TE polarization. The multipole decomposition of the modes can be calculated by integrating $\mathbf{P}$: \cite{Evlyukhin2013MultipoleSurface,Wu2020Room-TemperatureContinuum}


\begin{equation}
    \mathbf{p} = \int \mathbf{P}d\mathbf{r} \;,
\end{equation}

\begin{equation}
    \mathbf{m} = - \frac{i \omega}{2}\int (\mathbf{r}\times\mathbf{P})\mathbf{r}d\mathbf{r} \;,
\end{equation}

\begin{equation}
    \overline{\overline{\mathbf{Q}}} = 3 \int (\mathbf{rP} + \mathbf{Pr})d\mathbf{r} \;.
\end{equation}

The juxtaposed terms, such as $\mathbf{rP}$, represent the dyadic product of both vectors. The integrals are calculated over the volume of the scatterers (i.e., the Si nanodisks) using its geometric center as the coordinate origin. Note that on this basis, the electric quadrupole tensor is totally symmetric and not traceless, and the toroid-dipole terms would not appear. The contribution of the magnetic quadrupole and superior orders have been confirmed to be negligible. Once the multipolar terms are known, we can compute the scattering efficiency of each one with

\begin{equation}
    Q_{sca}^{ED_i} = \frac{k_0^4}{6\pi^2 \varepsilon_0^2 E_0^2 R^2} |p_i|^2
\end{equation}

\begin{equation}
    Q_{sca}^{MD_i} = \frac{\eta_0^2 \varepsilon_d k_0^4}{6\pi^2 E_0^2 R^2} |m_i|^2
\end{equation}

\begin{equation}
    Q_{sca}^{EQ_{ij}} = \frac{\varepsilon_d k_0^6}{80\pi^2 \varepsilon_0^2 E_0^2 R^2} |Q_{ij}|^2
\end{equation}

where $i, j = x,y,z$, $\eta_0$ is the vacuum impedance, and $R$ is the nano-disks' radius.

\begin{acknowledgement}
We thank Ramón Paniagua-Domínguez for sharing his COMSOL code to calculate the multipolar contributions to the resonances shown in Fig. 2. 
\end{acknowledgement}

\section*{Funding Sources}
This research is funded by the Innovational Research Incentives Scheme of the Nederlandse Organisatie voor Wetenschappelijk Onderzoek (NWO)  through  the  Gravitation  grant  “Research  Centre  for  Integrated  Nanophotonics”  and through the Vici Grant (680-47-628). It is also funded by the Spanish
Ministerio de Ciencia e Innovación through Grants MELODIA/PGC2018-095777-B-C21 and BICPLAN6G/TED2021-131417B-I00 (MCIN/AEI/10.13039/501100011033 and
European Union NextGenerationEU/PRTR), and by the University of Valladolid, within the Margarita Salas program, grant number CONVREC-2021-23 and by the Ministry of Education, Culture, Sports, Science and Technology (MEXT), Japan (22H01776), JSPS collaborative work (JPJSBP120219920), and the Asahi Glass Foundation.

\section*{Supporting Information}
The supporting information contains a simulation and measurements of the metasurface in a homogeneous environment, the mode decomposition of the four resonances. A description of the coupled oscillator model. The extinction of the metasurface under P-polarization, the visualization of the condensation in momentum ($k_x,k_y$) space and an analysis of the two BIC modes.

\bibliography{references}
\newpage

\end{document}